# A simple multithreaded implementation of the EM algorithm for mixture models

Sharon X. Lee[1], Kaleb L. Leemaqz[1], Geoffrey J. McLachlan[1,⋆]

[1]Department of Mathematics, University of Queensland, St. Lucia, Queensland, 4072, Australia.
⋆ E-mail: g.mclachlan@uq.edu.au

## Abstract

Finite mixture models have been widely used for the modelling and analysis of data from heterogeneous populations. Maximum likelihood estimation of the parameters is typically carried out via the Expectation-Maximization (EM) algorithm. The complexity of the implementation of the algorithm depends on the parametric distribution that is adopted as the component densities of the mixture model. In the case of the skew normal and skew $t$-distributions, for example, the E-step would involve complicated expressions that are computationally expensive to evaluate. This can become quite time-consuming for large and/or high-dimensional datasets. In this paper, we develop a multithreaded version of the EM algorithm for the fitting of finite mixture models. Due to the structure of the algorithm for these models, the E- and M-steps can be easily reformulated to be executed in parallel across multiple threads to take advantage of the processing power available in modern-day multicore machines. Our approach is simple and easy to implement, requiring only small changes to standard code. To illustrate the approach, we focus on a fairly general mixture model that includes as special or limiting cases some of the most commonly used mixture models including the normal, $t$-, skew normal, and skew $t$-mixture models.

## 1 Introduction

Finite mixture models are one of the most widely used tools for the modelling and analysis of heterogeneous data. They provide an elegant and flexible framework for handling heterogeneity within the data. Their usefulness has been demonstrated in numerous applications across a wide range of fields including astronomy, agriculture, biology, economics, engineering, finance, imaging, marketing, medicine, and social sciences. Comprehensive surveys on mixture models and their applications can be found in the books by Everitt and Hand (1981), Titterington et al. (1985), McLachlan and Basford (1988), Lindsay (1995), Böhning (2000), McLachlan and Peel (2000), and Frühwirth-Schnatter (2006), the edited volume of Mengersen et al. (2011), and also the papers by Banfield and Raftery (1993) and Fraley and Raftery (1998), among many others.

The normal distribution remains as the most popular choice of parametric families for the component densities of the mixture model. However, the use of non-traditional mixture models has become increasingly popular in recent years following the growing literature on non-normal and, in particular, skew distributions. These models are generalizations of the normal mixture model and provide very flexible distributional shapes with heavy tails and skewness. Some notable contributions include mixture modelling using, for example, skew elliptical component distributions (Wang et al. (2009), Lin (2010), Frühwirth-Schnatter and Pyne (2010), Cabral et al. (2012), Lin et al. (2014), Lee and McLachlan (2013, 2014, 2016)), multivariate normal inverse Gaussian component distributions (Karlis and Santourian (2009)), and multiple-scaled component distributions (Forbes and Wraith (2014), Wraith and Forbes (2015)).

Parameter estimation for finite mixture models can be carried out by maximum likelihood (ML) via the Expectation-Maximization (EM) algorithm (see the aforementioned references). However, the complexity of the algorithm depends on the chosen component densities. Whereas the E- and M-steps are rather straightforward for the case of the normal and $t$-mixture models, the expressions involved in these steps can be much more intricate for other mixture models. In the case of the skew normal mixture model, for example, the E-step would involve the calculation of the moments of the truncated normal distribution. Depending on the characterization of the (multivariate) skew normal distribution used, this requires the evaluations of multidimensional integrals that are computationally expensive to evaluate. Thus for high-dimensional and/or very large datasets, the computation cost of the algorithm can become very high. To reduce the computation time for fitting these models, we consider here a simple reformulation of the EM algorithm to take advantage of the multicore environments of modern computers.

In this paper, we introduce a straightforward approach to implement these EM algorithms in parallel. Due to the inherent structure of these algorithms, most of the expressions that need to be evaluated on the E- and M-steps can be performed independently for each individual component. Thus one may schedule these to be executed concurrently on different threads. This approach allows existing implementation to be easily adapted to support machines with multiple cores and processors, requiring minimal changes to existing code.

To illustrate our approach, we specialize the finite mixture model to the case of the multivariate skew $t$ (MST) mixture model since it includes the normal, $t$, and skew normal distributions as special and/or limiting cases. Also, we adopt the canonical fundamental skew $t$ (CFUST) distribution as component densities of our MST mixture model, a fairly general characterization that encompasses some of the more commonly used characterizations of the MST distribution including the classical formulation by Azzalini and Capitanio (2003) and the version by Sahu et al. (2003). For further details and discussions on the CFUST distribution and its link to various other versions of the MST distribution, readers are referred to the papers by Lee and McLachlan (2016) and McLachlan and Lee (2016), and the references therein. An EM algorithm for the fitting of finite mixtures of CFUST distributions was presented in the recent paper by Lee and McLachlan (2016) and its implementation in the R (R Core Team, 2016) was described in Lee and McLachlan (2015). Our approach for parallel implementation will be demonstrated with this EM algorithm.

The remainder of this paper is organised as follows. In Section 2, we briefly describe various commonly used mixture models, including the normal, $t$-, skew normal, and skew $t$-mixture models. The CFUST mixture model is then introduced as a generalization of these models. Section 3 provides an overview of the EM algorithm for fitting mixtures of CFUST distributions and its nested cases. In Section 4, we present the details of a multithreaded implementation of this algorithm. The methodology is demonstrated on some real datasets in Section 5. A summary is then given in Section 6.

## 2 Finite mixture models

A finite mixture model is a convex combination of component distributions. This provides a natural representation of data from heterogeneous populations, where each component of the mixture model is usually taken to correspond to the distribution of one of the populations. To establish notation, let $\boldsymbol{Y}$ be a $p$-dimensional random vector consisting of the $p$ feature variables of interest. We shall refer to $\boldsymbol{y}$ as a realization of $\boldsymbol{Y}$. The density for a $g$-component finite mixture model takes the form

$$f(\boldsymbol{y}; \boldsymbol{\Psi}) = \sum_{h=1}^{g} \pi_h \, f_h(\boldsymbol{y}; \boldsymbol{\theta}_h), \qquad (1)$$

where $\pi_h$ ($h = 1, \ldots, g$) are the mixing proportions and $f_h(\cdot)$ denotes the density of a mixture component. The mixing proportions are nonnegative and sum to one, that is, they satisfy $\pi_h \geq 0$ and $\sum_{h=1}^{g} \pi_h = 1$. The vector $\boldsymbol{\Psi} = (\pi_1, \ldots, \pi_{g-1}, \boldsymbol{\theta}_1^T, \ldots, \boldsymbol{\theta}_g^T)$ contains all the unknown parameters of the mixture model, with $\boldsymbol{\theta}_h$ containing the unknown parameters of the $h$th component. In most applications, the component densities $f_h(\cdot)$ are taken to be from the same parametric family. Traditionally, $f_h(\cdot)$ is the density of a multivariate normal distribution $\phi_p(\cdot)$ where $p$ denotes the dimension. Another popular choice for $f_h(\cdot)$ is a robust version of the normal distribution in the form of the multivariate $t$-distribution, denoted by $t_p(\cdot)$. The density of these two distributions are given by

$$\phi_p(\boldsymbol{y}; \boldsymbol{\mu}, \boldsymbol{\Sigma}) = (2\pi)^{-\frac{p}{2}} |\boldsymbol{\Sigma}|^{-\frac{1}{2}} e^{-\frac{1}{2}(\boldsymbol{y}-\boldsymbol{\mu})^T \boldsymbol{\Sigma}^{-1}(\boldsymbol{y}-\boldsymbol{\mu})} \qquad (2)$$

and

$$t_p(\boldsymbol{y}; \boldsymbol{\mu}, \boldsymbol{\Sigma}, \nu) = \frac{\Gamma(\frac{\nu+p}{2})}{\Gamma(\frac{\nu}{2})} (\nu p)^{-\frac{p}{2}} |\boldsymbol{\Sigma}|^{-\frac{1}{2}} \left[ 1 + \frac{(\boldsymbol{y}-\boldsymbol{\mu})^T \boldsymbol{\Sigma}^{-1}(\boldsymbol{y}-\boldsymbol{\mu})}{\nu} \right]^{(\frac{\nu+p}{2})}, \qquad (3)$$

respectively. In the above, $\boldsymbol{\mu}$ is a $p$-dimensional vector of location parameters, $\boldsymbol{\Sigma}$ is a positive-definite scale matrix, and $\nu$ is a scalar (called the degrees of freedom) that regulate the tails of the $t$-distribution.

A (multivariate) skew $t$ distribution is a generalization of the $t$-distribution. It has extra parameters for the regulation of the skewness of the distribution. There are various characterizations of the MST distribution which correspond to different mechanisms used to introduce skewness to the $t$-distribution. The canonical fundamental skew $t$-distribution (CFUST) distribution was introduced as a member of the family of fundamental skew distribution by Arellano-Valle and Genton (2005). This is a fairy general characterization of the skew distribution that encompasses many other existing formulations of skew distributions. The skewness parameters of the CFUST distribution take the form of a $p \times q$ matrix denoted by $\boldsymbol{\Delta}$. Its density is given by

$$f_{\text{CFUST}}(\boldsymbol{y}; \boldsymbol{\mu}, \boldsymbol{\Sigma}, \boldsymbol{\Delta}, \nu) = 2^q \, t_p(\boldsymbol{y}; \boldsymbol{\mu}, \boldsymbol{\Omega}, \nu) \, T_q\left( \boldsymbol{c}(\boldsymbol{y}) \sqrt{\frac{\nu+p}{\nu+d(\boldsymbol{y})}}; \boldsymbol{0}, \boldsymbol{\Lambda}, \nu+p \right), \qquad (4)$$

where

$$\begin{aligned}
\boldsymbol{\Omega} &= \boldsymbol{\Sigma} + \boldsymbol{\Delta}\boldsymbol{\Delta}^\top, \\
\boldsymbol{c}(\boldsymbol{y}) &= \boldsymbol{\Delta}^\top \boldsymbol{\Omega}^{-1} (\boldsymbol{y} - \boldsymbol{\mu}), \\
\boldsymbol{\Lambda} &= \boldsymbol{I}_q - \boldsymbol{\Delta}^\top \boldsymbol{\Omega}^{-1} \boldsymbol{\Delta}, \\
d(\boldsymbol{y}) &= (\boldsymbol{y} - \boldsymbol{\mu})^\top \boldsymbol{\Omega}^{-1} (\boldsymbol{y} - \boldsymbol{\mu}).
\end{aligned}$$

In the above, we let $t_q(\boldsymbol{y}; \boldsymbol{\mu}, \boldsymbol{\Omega}, \nu)$ denote the $q$-dimensional $t$-distribution as defined by (3) and $T_q(.)$ is its corresponding (cumulative) $t$-distribution function. It can be observed from (4) that the CFUST distribution includes the multivariate normal and $t$-distributions as special and/or limiting cases. More specifically, taking $\boldsymbol{\Delta} = \boldsymbol{0}$ reduces (4) to the symmetric multivariate $t$-density (3), and further letting $\nu \to \infty$ leads to the multivariate normal density (2).

Another important limiting case of the MST distribution is the multivariate skew normal (MSN) distribution, which is obtained by letting $\nu \to \infty$. For the CFUST characterization of the MST distribution, the corresponding MSN distribution is the canonical fundamental skew normal (CFUSN) distribution with density given by

$$f_{\text{CFUSN}}(\boldsymbol{y}; \boldsymbol{\mu}, \boldsymbol{\Sigma}, \boldsymbol{\Delta}) = 2^q \phi_p(\boldsymbol{y}; \boldsymbol{\mu}, \boldsymbol{\Omega}) \; \Phi_q(\boldsymbol{c}(\boldsymbol{y}); \boldsymbol{0}, \boldsymbol{\Lambda}). \tag{5}$$

As mentioned previously, the CFUST distribution encompasses some of the existing characterizations of the MST and the MSN distributions. To note some of the more commonly used ones here, if $\boldsymbol{\Delta}$ is constrained to be a diagonal matrix, then we obtain the skew $t$-distribution of Sahu et al. (2003) which is referred to as the unrestricted skew $t$-distribution using the terminology in Lee and McLachlan (2014). To obtain the classical skew $t$-distribution by Azzalini and Capitanio (2003) from (4), one can set $q = 1$ or take $\boldsymbol{\Delta}$ to be a matrix of zeros except for one column (Lee and McLachlan, 2016). This formulation of the skew $t$-distribution, referred to as the restricted skew $t$-distribution by Lee and McLachlan (2013), is equivalent to that given by Branco and Dey (2001), Gupta (2003), Lachos et al. (2010) and Pyne et al. (2009). The restricted skew $t$-distribution is restricted in the sense that it is limited to modelling skewness concentrated in one direction in the feature space (McLachlan and Lee, 2016). Analogous formulations exist for the CFUSN distribution; that is, the restricted and unrestricted skew normal distributions can be obtained by placing appropriate constraints on $\boldsymbol{\Delta}$ and letting $\nu \to \infty$ in (5).

In this sequel, we shall work with the skew $t$-mixture model with components taken to belong to the CFUST family of distributions.

## 3 Parameter estimation via the EM algorithm for the CFUST mixture model

Before we proceed to present a multithreaded EM algorithm for the ML likelihood estimation of the parameters of the skew $t$-mixture model with CFUST component distributions, we first describe the EM algorithm implemented in standard form for the iterative computation of the ML estimates.

To facilitate parameter estimation for mixture models via the EM algorithm, a set of latent variables are introduced, namely the component label $\boldsymbol{Z}_j$, corresponding to the $j$th observation $\boldsymbol{y}_j$ ($j = 1, \ldots, n$) in the conceptual EM framework in which $\boldsymbol{Y}_j$ belongs to one of the $g$ components in the mixture model. Here $\boldsymbol{Z}_j$ is a $g$-dimensional vector of binary component labels such that $Z_{hj} = (\boldsymbol{Z}_j)_h = 1$ if the $j$th observation belongs to the $h$th component and zero otherwise. In the case of non-normal distributions, other latent variables may also be introduced. For instance, latent gamma variables $W_j$ are introduced for the $t$-mixture model that relates to its degrees of freedom. For skew distributions, another set of latent variables $\boldsymbol{U}_j$ that relates to the skewness of the distribution is needed in its convolution-type stochastic representation. In the case of the FM-CFUST model, we would have the $\boldsymbol{Z}_j$, $W_j$, and $\boldsymbol{U}_j$ as missing variables, where the latter follows a half-normal distribution (conditional on $W_j$). Thus, the complete-data for the FM-CFUST model consist of these missing variables and the observations $\boldsymbol{y}_j$. This leads to a four-level hierarchical characterization of the FM-CFUST model, given by

$$
\begin{aligned}
\boldsymbol{Y}_j \mid \boldsymbol{u}_j, w_j, z_{hj} = 1 &\sim N_p\left(\boldsymbol{\mu} + \boldsymbol{\Delta}_h \boldsymbol{u}_j, \tfrac{1}{w_j}\boldsymbol{\Sigma}_h\right), \\
\boldsymbol{U}_j \mid w_j, z_{hj} = 1 &\sim HN_q\left(\boldsymbol{0}, \tfrac{1}{w_j}\boldsymbol{I}_q\right), \\
W_j \mid z_{hj} = 1 &\sim \mathrm{gamma}\left(\tfrac{\nu_h}{2}, \tfrac{\nu_h}{2}\right), \\
\boldsymbol{Z}_j &\sim \mathrm{Mult}_g(1, \boldsymbol{\pi}).
\end{aligned} \quad (6)
$$

Here $HN_q(\boldsymbol{0}, \boldsymbol{\Sigma})$ denotes the $q$-dimensional half-normal distribution with scale matrix $\boldsymbol{\Sigma}$, gamma$(\alpha, \beta)$ denotes the gamma distribution with mean $\alpha/\beta$, and $\mathrm{Mult}_g(1, \boldsymbol{\pi})$ denotes the multinomial distribution with one draw and $g$ categories with probabilities specified by $\boldsymbol{\pi}$.

## 3.1 E-step

The E-step of the EM algorithm requires the calculation of the so-called $Q$-function, $Q(\boldsymbol{\Psi}; \boldsymbol{\Psi}^{(k)})$, which is the conditional expectation of the complete-data log likelihood given the observed data $\boldsymbol{y}$, using the current estimate of $\boldsymbol{\Psi}$, which is denoted by $\boldsymbol{\Psi}^{(k)}$ after the $k$th iteration. It follows that on the $(k+1)$th iteration, the E-step requires the following five conditional expectations to be calculated,

$$
\begin{aligned}
z_{hj}^{(k)} &= E_{\Psi^{(k)}}\left[z_{hj} = 1 \mid \boldsymbol{y}_j\right], & (7) \\
w_{hj}^{(k)} &= E_{\Psi^{(k)}}\left[w_{hj} \mid \boldsymbol{y}_j, z_{hj} = 1\right], & (8) \\
e_{1hj}^{(k)} &= E_{\Psi^{(k)}}\left[\log(w_{hj}) \mid \boldsymbol{y}_j, z_{hj} = 1\right], & (9) \\
\boldsymbol{e}_{2hj}^{(k)} &= E_{\Psi^{(k)}}\left[w_{hj}\boldsymbol{u}_{hj} \mid \boldsymbol{y}_j, z_{hj} = 1\right], & (10) \\
\boldsymbol{e}_{3hj}^{(k)} &= E_{\Psi^{(k)}}\left[w_{hj}\boldsymbol{u}_{hj}\boldsymbol{u}_{hj}^\top \mid \boldsymbol{y}_j, z_{hj} = 1\right]. & (11)
\end{aligned}
$$

These expectations can be expressed as

$$
z_{hj}^{(k)} = \frac{\pi_h f(\boldsymbol{y}_j; \boldsymbol{\mu}_h^{(k)}, \boldsymbol{\Sigma}_h^{(k)}, \boldsymbol{\delta}_h^{(k)}, \nu_h^{(k)})}{f(\boldsymbol{y}_j; \boldsymbol{\Psi}^{(k)})}, \quad (12)
$$

$$
w_{hj}^{(k)} = \left(\frac{\nu_h^{(k)} + p}{\nu_h^{(k)} + d_h^{(k)}(\boldsymbol{y}_j)}\right) \frac{T_q\left(\boldsymbol{c}_{hj}^{(k)} \sqrt{\frac{\nu_h^{(k)}+p+2}{\nu_h^{(k)} d_h^{(k)}(\boldsymbol{y}_j)}}; \boldsymbol{0}, \boldsymbol{\Lambda}_h^{(k)}, \nu_h^{(k)} + p + 2\right)}{T_q\left(\boldsymbol{c}_{hj}^{(k)} \sqrt{\frac{\nu_h^{(k)}+p}{\nu_h^{(k)} d_h^{(k)}(\boldsymbol{y}_j)}}; \boldsymbol{0}, \boldsymbol{\Lambda}_h^{(k)}, \nu_h^{(k)} + p\right)}, \quad (13)
$$

$$
(14)
$$

$$e_{1hj}^{(k)} = w_{hj}^{(k)} - \log\left(\frac{\nu_h^{(k)} + d_h^{(k)}(\boldsymbol{y}_j)}{2}\right) - \left(\frac{\nu_h^{(k)} + p}{\nu_h^{(k)} + d_h^{(k)}(\boldsymbol{y}_j)}\right) + \psi\left(\frac{\nu_h^{(k)} + p}{2}\right), \quad (15)$$

$$\boldsymbol{e}_{2,hj}^{(k)} = w_{hj}^{(k)} E_{\boldsymbol{\Psi}^{(k)}}[\boldsymbol{u}_{hj} \mid \boldsymbol{y}_j], \quad (16)$$

$$\boldsymbol{e}_{3hj}^{(k)} = w_{hj}^{(k)} E_{\boldsymbol{\Psi}^{(k)}}[\boldsymbol{u}_{hj}\boldsymbol{u}_{hj}^T \mid \boldsymbol{y}_j], \quad (17)$$

where $\boldsymbol{U}_{hj} \mid \boldsymbol{y}_j$ has a $q$-dimensional truncated $t$-distribution given by

$$\boldsymbol{U}_{hj} \mid \boldsymbol{y}_j \sim tt_q\left(\boldsymbol{c}_{hj}^{(k)}, \left(\frac{\nu_h^{(k)} + d_h(\boldsymbol{y}_j)}{\nu_h^{(k)} + p + 2}\right)\boldsymbol{\Lambda}_h^{(k)}, \nu_h^{(k)} + p + 2; \mathbb{R}^+\right).$$

It can be observed from (16) and (17) above that these two conditional expectations are the first and second moments of $\boldsymbol{U}_{hj}$. It is also of interest to note that $e_{1hj}^{(k)}$ can be evaluated using different approaches as described in Lee and McLachlan (2015). Here we adopt the one-step-late (OSL) approach as implemented in the EMMIX packages.

## 3.2 M-step

The M-step is relatively straightforward compared to the E-step. On the $(k+1)$th iteration of the M-step, the current estimate of $\boldsymbol{\Psi}$, $\boldsymbol{\Psi}^{(k)}$, is updated to $\boldsymbol{\Psi}^{(k+1)}$, which is chosen to globally maximize $Q(\boldsymbol{\Psi}; \boldsymbol{\Psi}^{(k)})$ over $\boldsymbol{\Psi}$. For the FM-CFUST model, the M-step leads to the following updates:

$$\pi_h^{(k+1)} = \frac{1}{n}\sum_{j=1}^n z_{hj}^{(k)}, \quad (18)$$

$$\boldsymbol{\mu}_h^{(k+1)} = \frac{\sum_{j=1}^n z_{hj} w_{hj}^{(k)} \boldsymbol{y}_j - \boldsymbol{\Delta}_h^{(k)} \sum_{j=1}^n z_{hj}^{(k)} \boldsymbol{e}_{2hj}^{(k)}}{\sum_{j=1^n}^n z_{hj}^{(k)} w_{hj}^{(k)}}, \quad (19)$$

$$\boldsymbol{\Delta}_h^{(k+1)} = \left[\sum_{j=1}^n z_{hj}^{(k)}\left(\boldsymbol{y}_j - \boldsymbol{\mu}_h^{(k+1)}\right)\boldsymbol{e}_{2hj}^{(k)\top}\right]\left[\sum_{j=1}^n z_{hj}^{(k)} \boldsymbol{e}_{3hj}^{(k)}\right]^{-1}, \quad (20)$$

$$\boldsymbol{\Sigma}_h^{(k+1)} = \left\{\sum_{j=1}^n z_{hj}^{(k)}\left[w_{hj}^{(k)}\left(\boldsymbol{y}_j - \boldsymbol{\mu}_h^{(k+1)}\right)\left(\boldsymbol{y}_j - \boldsymbol{\mu}_h^{(k+1)}\right)^T - \boldsymbol{\Delta}_h^{(k+1)} \boldsymbol{e}_{2hj}^{(k)}\left(\boldsymbol{y}_j - \boldsymbol{\mu}_h^{(k+1)}\right)^\top\right.\right.$$
$$\left.\left. - \left(\boldsymbol{y}_j - \boldsymbol{\mu}_h^{(k+1)}\right)\boldsymbol{e}_{2hj}^{(k)\top}\boldsymbol{\Delta}_h^{(k+1)\top} + \boldsymbol{\Delta}_h^{(k+1)}\boldsymbol{e}_{3hj}^{(k)\top}\boldsymbol{\Delta}_h^{(k+1)\top}\right]\right\}\left[\sum_{j=1}^n z_{hj}^{(k)}\right]^{-1}. \quad (21)$$

An update of the degrees of freedom $\nu_h$ is obtained by solving the following equation for $\nu_h^{(k+1)}$,

$$0 = \left(\sum_{h=1}^n z_{hj}^{(k)}\right)\left[\log\left(\frac{\nu_h^{(k+1)}}{2}\right) - \psi\left(\frac{\nu_h^{(k+1)}}{2}\right) + 1\right] - \sum_{j=1}^n z_{hj}^{(k)}\left(e_{1hj}^{(k)} - w_{hj}^{(k)}\right), \quad (22)$$

where $\psi(\cdot)$ denotes the digamma function.

## 3.3 Initialization

As with the case of traditional mixture models, the EM algorithm described above is sensitive to its initialization and so it is important to choose good starting values for the parameters of the model. Some strategies for finding sensible initial values for the FM-CFUST model are presented in Lee and McLachlan (2015). We adopt the moments-based approach described there. It leads to the following set of initial values,

$$\begin{aligned}
\boldsymbol{\delta}^{(0)} &= \pm\sqrt{\frac{\pi(1-\alpha)}{\pi-2}}\boldsymbol{s}^*, \\
\boldsymbol{\Sigma}^{(0)} &= \boldsymbol{S} - (1-\alpha)\operatorname{diag}(\boldsymbol{s}^*), \\
\boldsymbol{\mu}^{(0)} &= \bar{\boldsymbol{y}} - \sqrt{\frac{2}{\pi}}\boldsymbol{\delta}^{(0)},
\end{aligned} \quad (23)$$

where $\bar{\boldsymbol{y}}$ denotes the sample mean, $\boldsymbol{s}^*$ is the vector containing the diagonal elements of the sample covariance matrix $\boldsymbol{S}$, and $\alpha$ is an arbitrary value in $(0, 1)$. The sign of each element of $\boldsymbol{\delta}^{(0)}$ is taken to be the sign of the third-order sample moment of the corresponding variable about its sample mean.

Given an initial partition of the data, a set of initial values for the CFUST mixture model can be obtained by applying the specification (23) to each cluster individually. This is repeated over a range of values of $\alpha$ and the set of initial parameters corresponding to the highest log likelihood value is chosen. In practice, multiple different initial clusterings are generated, typically based on $k$-means or random starts, and then the above procedure is performed on each of these clusterings. Their log likelihood values are compared and the best performing set of initial values are used to start the EM iterations.

In summary, the initialization process proceeds as shown below in Algorithm 1.

---

**Algorithm 1** Initialization of EM algorithm

1: **procedure** INITIALIZATION
2:     $l \leftarrow -\infty$
3:     $\boldsymbol{\Psi} = \emptyset$
4:     **for** i=1 to nStarts **do**
5:         Generate a new partition of the data
6:         Set $\pi_1, \pi_2, \ldots, \pi_g$ according to the partition
7:         **for** a=0.1 to 0.9 **do**
8:             **for** h=1 to g **do**
9:                 Compute $\bar{\boldsymbol{y}}$, $\boldsymbol{S}$, and $\boldsymbol{s}^*$ for cluster $h$
10:                Set $\boldsymbol{\mu}_h, \boldsymbol{\Sigma}_h, \boldsymbol{\Delta}_h$, and $\nu_h$ using (23)
11:                $\boldsymbol{\theta}_h \leftarrow \{\boldsymbol{\mu}_h, \boldsymbol{\Sigma}_h, \boldsymbol{\Delta}_h, \nu_h\}$
12:             $l^* \leftarrow$ log-likelihood value
13:             **if** $l^* > l$ **then**
14:                $\boldsymbol{\Psi} \leftarrow \{\boldsymbol{\theta}_1, \boldsymbol{\theta}_2, \ldots, \boldsymbol{\theta}_g, \pi_1, \ldots, \pi_{g-1}\}$
    **return** $\boldsymbol{\Psi}$

## 3.4 Check for convergence

Following Lee and McLachlan (2016), the Aitken acceleration-based stopping criterion is adopted to assess the convergence of the EM algorithm. Briefly, we terminate the EM algorithm when the absolute difference between a log likelihood value and its asymptotic estimate is smaller than the tolerance $\epsilon = 10^{-6}$. More formally, we stop at the $(k+1)$th iteration when

$$\left|\ell_\infty^{(k+1)} - \ell^{(k+1)}\right| < \epsilon, \tag{24}$$

where $\ell^{(k+1)}$ is the log likelihood value at the $(k+1)$th iteration and $\ell_\infty^{(k+1)}$ is its asymptotic estimate. The latter involves $a^{(k)}$ which is Aitken's acceleration at the $k$th iteration. These two quantities are defined as

$$\ell_\infty^{(k+1)} = \ell^{(k)} + \frac{\ell^{(k+1)} - \ell^{(k)}}{1 - a^{(k)}}, \tag{25}$$

and

$$a^{(k)} = \frac{\ell^{(k+1)} - \ell^{(k)}}{\ell^{(k)} - \ell^{(k-1)}}. \tag{26}$$

Existing implementation of the above EM algorithm performs the initialization, EM-steps, and the check for convergence in a serial manner.

## 4 A multithreaded implementation of the EM algorithm

As mentioned previously, the computation of the moments (16) and (17) in the E-step can become very time-consuming when the number of variables $p$ or, more precisely, the number of latent skewing variables $q$, is large. To reduce the computation time required for the fitting of the FM-CFUST model, we consider the approach of parallel programming that is being used more frequently these days with the increasing need to analyse big datasets.

Most modern machines offer a multi-core environment for improved performance. This enables multiple threads to be run simultaneously. The time performance of algorithms such as the EM algorithm described in this paper can benefit substantially if they can be reformulated to be used on machines with multiple cores of CPUs. To take advantage of these computing resources, we developed a multithreaded implementation of the EM algorithm described in the previous section. To our knowledge, there is no available parallel implementation for the fitting of finite mixtures of skew distributions in the literature. There are, however, some proposals for parallelizing the EM algorithm for normal mixture models; see for example, Kwedlo (2014), Kapoor et al. (2015), Kumar et al. (2009), Guo et al. (2012), and Chen et al. (2013), and the references therein.

### 4.1 Parallel E-step

Due to the structure of the EM algorithm for the fitting of finite mixture models, most of the conditional expectations on the E-steps can be performed independently for each component $h$. As can be observed from (13) to (17) for the case of the FM-CFUST model, these four conditional expectations for component $h$ do not involve the parameters or conditional expectations from other components. This means that $w_{hj}^{(k)}$, $e_{1,hj}^{(k)}$, $\boldsymbol{e}_{2,hj}^{(k)}$, and $\boldsymbol{e}_{3,hj}^{(k)}$ can be performed simultaneously on different threads. These sets of computations are independent and do not require communication between the threads. Concerning the first conditional expectation $z_{hj}^{(k)}$, the expression (12) indicates that parameters from other components are required only in the evaluation of the denominator of this expression. However, the latter quantity is given by the sum of the numerator across all $g$ components for the $j$th observation. Moreover, it can observed that the $z_{hj}^{(k)}$ are not used elsewhere on the E-step. Hence one can split up the calculation of the $z_{hj}^{(k)}$ by allowing their numerator to be evaluated individually for each component $h$ and together with the remaining four conditional expectations. These results for the numerators are combined latter to form the denominator and to obtain the estimate $z_{hj}^{(k)}$.

It follows that the input for thread $h$ is the data $\boldsymbol{y}$ and the current estimate of the parameters for component $h$ of the model. At the end of the process, we collect from thread $h$ the values of the conditional expectations ($w_{hj}^{(k)}$, $e_{1,hj}^{(k)}$, $\boldsymbol{e}_{2,hj}^{(k)}$, and $\boldsymbol{e}_{3,hj}^{(k)}$) and the numerator of $z_{hj}^{(k)}$. This leads to the following set of tasks for thread $h$:

1. Calculate the numerator of $z_{hj}^{(k)}$ as defined (12) for $j = 1, \ldots, n$.

2. Calculate $w_{hj}^{(k)}$ as defined in (13) for $j = 1, \ldots, n$.

3. Calculate $e_{1,hj}^{(k)}$ as defined in (15) for $j = 1, \ldots, n$.

4. Calculate $\boldsymbol{e}_{2,hj}^{(k)}$ as defined in (16) for $j = 1, \ldots, n$.

5. Calculate $\boldsymbol{e}_{3,hj}^{(k)}$ as defined in (17) for $j = 1, \ldots, n$.

6. Return the above values to the master thread.

## 4.2 Parallel M-step

From Section 3.2, it can be observed that the M-step is easily parallizable across the $g$ components. All of the expressions for component $h$ do not require any knowledge of the parameters or conditional expectations from other components. Hence it is straightforward to split the calculations across the $g$ threads. It follows that the following set of tasks can be implemented for thread $h$:

1. Calculate $\pi_h^{(k)}$ as defined in (18).

2. Calculate $\boldsymbol{\mu}_h^{(k)}$ as defined in (19).

3. Calculate $\boldsymbol{\Sigma}_h^{(k)}$ as defined in (21).

4. Calculate $\boldsymbol{\Delta}_h^{(k)}$ as defined in (20).

5. Calculate $\nu_h^{(k)}$ using (22).

6. Return the above values to the master thread.

## 4.3 Job scheduling in the EM loops

It follows from from (18) and the previous section that the calculation of $\pi_h^{(k)}$ requires $z_{hj}^{(k)}$ to be calculated for $j = 1, \ldots, n$. Hence the results from Step 6 of the E-step need to be collected from the threads and combined to obtain the $z_{hj}^{(k)}$ before starting the M-step. This implies a 'collection' step is required before the M-step.

Furthermore, recall that at the end of each EM iteration, a check for convergence is performed to determine whether the algorithm should be stopped. To proceed, the current value of the log likelihood function needs to be computed. This is given by the sum of the logarithm of the density of the FM-CFUST model evaluated at the data points; that is, it is given by

$$\ell^{(k)} = \sum_{j=1}^{n} \log f(\boldsymbol{y}_j; \boldsymbol{\Psi}). \tag{27}$$

It can be observed from (27) that the terms in the summation are the same as the denominator of $z_{hj}^{(k+1)}$. Thus one can save time by not recomputing $z_{hj}^{(k)}$ again in the E-step of the next iteration. In other words, Step 1 of the E-step described in Section 4.1 can be performed on the M-step of the same iteration, after the computation of the updated estimates of the parameters of the model. After this, the numerators of the $z_{hj}^{(k)}$ as returned by the threads can be combined to obtain both the log likelihood value $\ell^{(k)}$ and the $z_{hj}^{(k+1)}$. Note that with this approach, results of the E-step need not be collected before the M-step. It follows that the M-step for a component can be performed straight after the E-step of the corresponding component on the same thread without interruption, and the collection of results from different threads occurs only at the end of an iteration.

## 5 Applications

For illustration, the algorithm is applied to two real datasets. Our first dataset concerns measurements from a blood sample collected by the British Columbia Cancer Agency. It consists of the fluorescence intensities of over 6000 cells from a blood sample donated by patient of a hematopoietic stem cell transplant (HSCT) experiment. The four variables correspond to four different protein markers used by experts to discriminate between four cell populations in the sample. A three-dimensional plot of the data is shown in the left panel of Figure 1, where the colours of the cells correspond to the labels given by expert analysts using manual gating.

We consider also the well-known Iris data (Fisher, 1936) originally collected by Anderson (1935). The 150 observations in the data consist of 50 samples from each of the three species of Iris, namely, Setosa, Versicolor, and Virginica. Four geometric measurements were taken from each sample, which are the length and width of the sepal and petal.

For these two datasets, the number of components $g$ is taken to be fixed since we know the true number of groups. The FM-CFUST model was first fitted using the algorithm described in Section 3. The same model was fitted again using the multicore implementation described in Section 4. Their performance in computation time is then discussed.

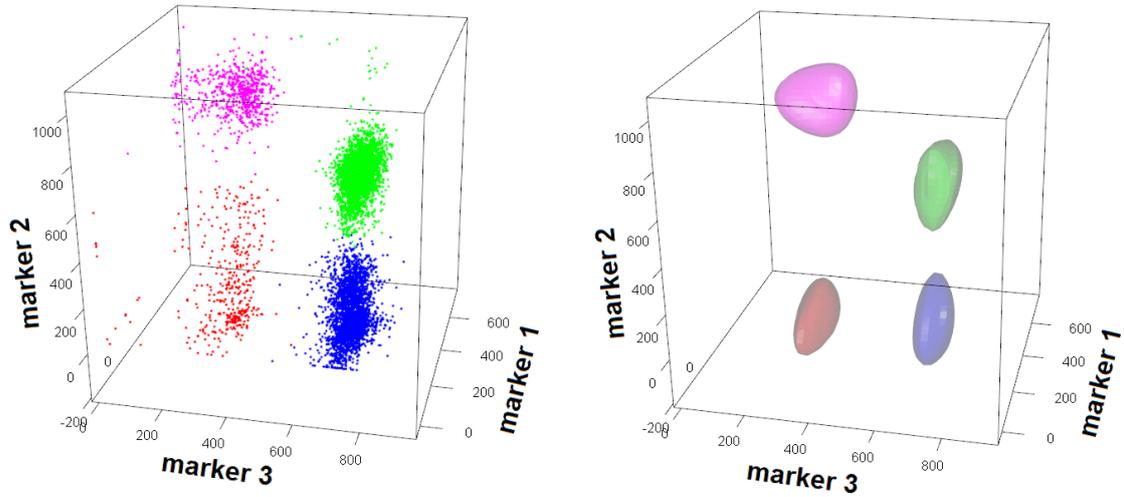

Figure 1: A scatter plot of the data in 3D is shown in the left panel and the density contours of each component of the fitted FM-CFUST model is shown in the right panels.

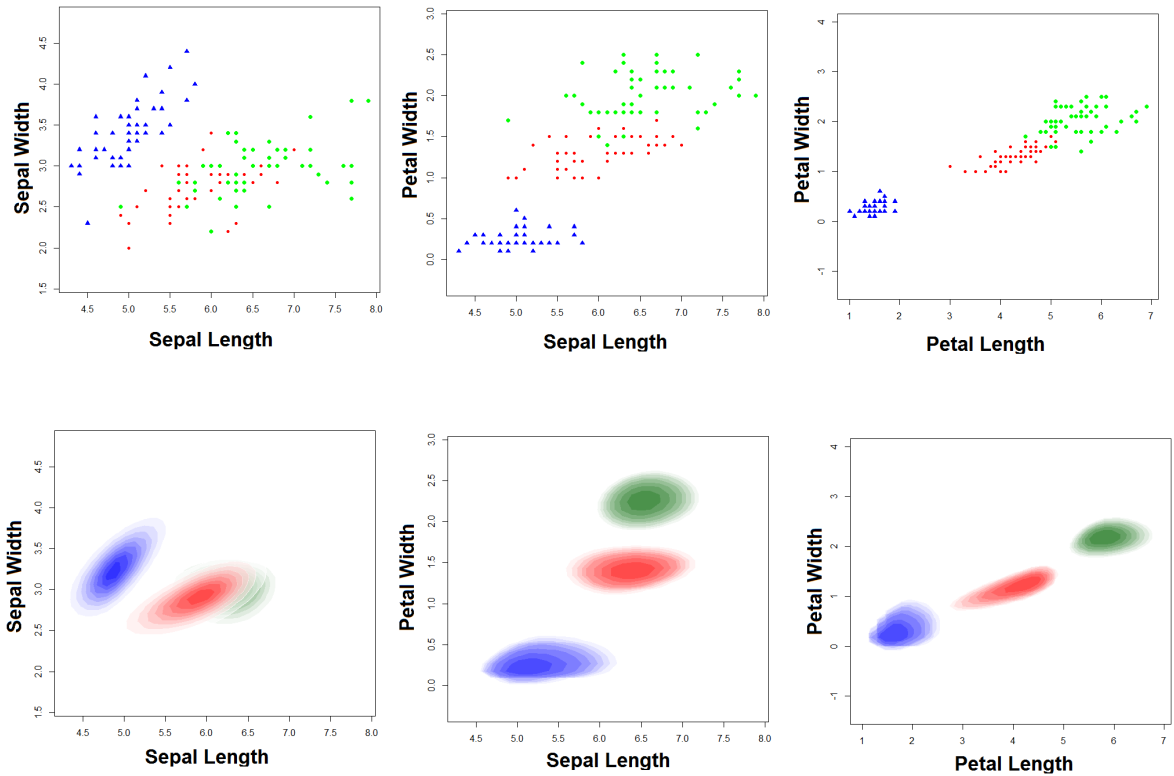

Figure 2: Bivariate plots of different combinations of variables of the Iris dataset. Top row: Scatter plots showing the true labels of the Iris dataset, where the Iris species Setosa, Versicolor and Virginica are shown in blue, red, and green, respectively. Bottom row: The density contours of each component of the fitted FM-CFUST model.

|           | MCR     | ARI     | $F$-measure | Time Reduction (%) |
|-----------|---------|---------|-------------|--------------------|
| FM-CFUST  | 0.00277 | 0.99305 | 0.9972      | 60.40              |
| FM-rMST   | 0.00358 | 0.99274 | 0.9964      | 58.81              |
| FM-CFUSN  | 0.00407 | 0.99124 | 0.9959      | 62.48              |

Table 1: Clustering and time performance of the FM-CFUST, FM-rMST, and FM-CFUSN models on the HSCT dataset. Columns two to four shows the MCR, ARI, and $F$-measure values, respectively. Column five indicates the reduction in time (in percentage) achieved using the multithreaded implementation.

## 5.1 Clustering performance

To assess the performance of the FM-CFUST model for the two datasets, we calculated its misclassification rate (MCR) against the true labels. For the HSCT dataset, the true labels are taken to be the labels given by experts, whereas for the Iris dataset they are given by the real species category of Iris. The MCR measure gives the proportion of misclassified observations, calculated by choosing among all possible permutations of the cluster labels the one that gives the lowest value.

The contours of the densities of the fitted FM-CFUST model for the HSCT dataset as depicted in the right panel of Figure 1 would appear to provide a reasonably close fit to the data shown in the left panel. The shape of the pink, green, and blue clusters seem to be captured reasonably well by the contours of the corresponding components of the fitted FM-CFUST model. For comparison, the FM-CFUSN and FM-rMST models were also fitted. The former attained a MCR of 0.00407 for this dataset, whereas the FM-CFUST model reduced it by around 32% (MCR = 0.00277). The clustering performance of the FM-rMST model is better than the FM-CFUSN model but not the FM-CFUST in this case, attaining a MCR between that of the FM-CFUSN and FM-CFUST models (MCR=0.00358). The corresponding adjusted Rand index (ARI) (Hubert and Arabie, 1985) and $F$-measure (Aghaeepour et al., 2013) are reported in the third and fourth columns of Table 1. These two measures provide a quantitative index for the agreement between two sets of labels for class memberships. An ARI value of one indicates perfect match with the true labels. The $F$-measure, defined as the harmonic mean of precision and recall, is perhaps a more commonly used measure to compare the performance of algorithms in flow cytometric analysis. Similar to the ARI, a higher $F$-measure value indicates a relatively better match with the true labels and a value of one implies a perfect match.

For the Iris dataset, the FM-CFUST model attained a MCR of 0.00667 and an ARI of 0.980, which implies only one misallocation. This is in agreement with the results reported in Lee and McLachlan (2016). They have also reported the performance of some other mixture models (including normal, $t$, and generalized hyperbolic mixture models), with MCR ranging from 0.0200 to 0.1067 and ARI values ranging from 0.732 to 0.941. The contours of the densities of the fitted FM-CFUST model for this dataset are depicted in the bottom row of Figure 2, showing a reasonable close match to the distribution of the data.

## 5.2  Reduction in computation time

As our algorithm is implemented by splitting up the calculations across $g$ threads, it would be expected that the potential performance gain will increase as the number of components $g$ increases. However, the increase in performance gain is not expected to be linear due to the overheads associated with the setting up of the parallel process. By running the multithreaded version of the implementation of the EM algorithm, we observed a reduction of (total) computation time by approximately 60% for the HSCT dataset using a typical quad-core machine. A similar reduction in time was also observed for the FM-rMST and FM-CFUSN models, which is about 62% and 58% respectively (see Table 1). For the Iris dataset, it was observed that the multithreaded version of the implementation of the EM algorithm reduces the computation time by around 48.9% for the FM-CFUST model.

Figure 3 shows the reduction in computation time for each iteration of the EM algorithm when applied to these two datasets. In the top panel of Figure 3, the time reduction for the FM-CFUST model is shown in red, whereas that for the FM-CFUSN and FM-rMST models was shown in blue and black, respectively. As can be expected, it can observed from the top panel of Figure 3 that the percentage of time reduction for the HSCT dataset is somewhat similar across the iterations, with the curve for the FM-CFUSN model slightly increasing for the later iterations. For the Iris dataset shown in the bottom panel of Figure 3, the percentage reduction in time for the FM-CFUST model appear to quite consistent across the iterations.

We have also implemented a specialized version for the nested case of the normal mixture model. However, the performance gain may not be as rewarding as in the case of the FM-CFUST model and may indeed be slower for small values of $g$, due to the relative overheads involved. However, the performance gain would still be expected to improve as $g$ increases. We have observed in our experiment on the Protein dataset from the 2004 KDD Cup that a reduction in time of approximately 30% to 50% was achieved for the normal mixture model for $g$ ranging from 2 to 20. It is apparent that our approach would be best suited to mixture models where the E and/or M-steps are computationally expensive. For instance, the skew-normal and skew $t$-mixture models would benefit more from this multithreaded implementation than the traditional normal and $t$-mixture models since their E-steps involve heavy calculations.

## 6  Conclusions

We have presented a straightforward parallelization scheme for implementing the EM algorithm for the fitting of finite mixtures models. Our approach is easy to implement and can be applied to the EM algorithm for any finite mixture model. To illustrate our approach, we considered only the case of a finite mixture of CFUST distributions, but it includes the normal, $t$, and skew normal mixture models as special and/or limiting cases. Also, with the CFUST mixture model, there is much potential available for reducing the computational time of the EM algorithm.

The performance gain of our approach was demonstrated on two real datasets. In the first datset which involves $g = 4$ components, it was found that a 60% reduction in computation time was obtained. For the second dataset with $g = 3$ components, a 50% reduction in computation time was achieved. In these two datasets, the number of threads used was equal to the number of components of the fitted model, which is four and three, respectively.

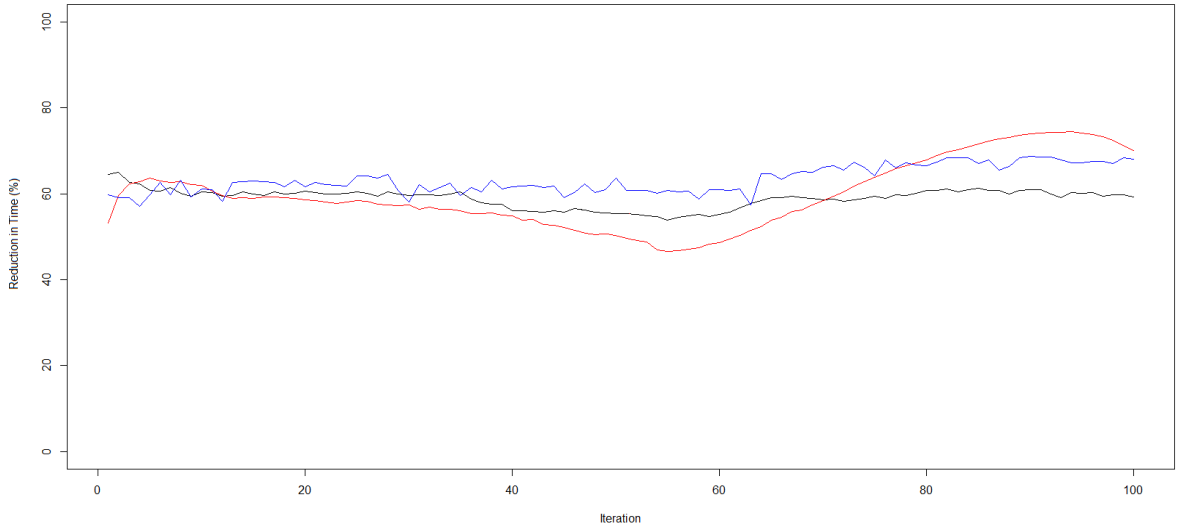

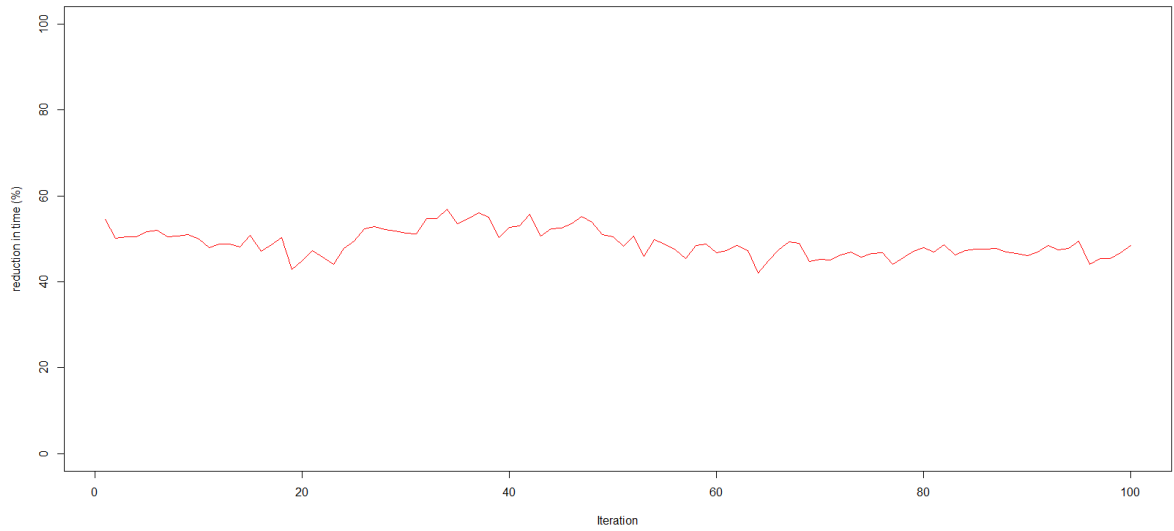

Figure 3: Top panel: Observed reduction in computation time (in percentage) by using the multithreaed version of the implementation of the EM algorithm. This is shown for each iteration of the EM algorithm for the FM-CFUST model (red line), FM-CFUSN model (blue line), and FM-rMST model (black line) on the HSCT dataset. Bottom panel: The percentage reduction in computation time for each iteration of the EM algorithm for the FM-CFUST model on the Iris dataset.

We have also carried out some experiments with the specialised case of a normal mixture model. However, the performance gain is not expected to be as rewarding as for the FM-CFUST model since the former does not involves time-consuming calculations in the E-step. We found that the reduction in time for a large dataset ($n = 145,751$ and $p = 40$) ranged between 30% to 60% for the cases where $g = 2, \ldots, 20$. As can be observed from the results in Section 5.2, our approach is more suited to skew mixture models such as the CFUSN and CFUST mixture models where computation cost is high in the standard implementation of the EM algorithm. Consequently, an appreciably higher reduction in computation time can be achieved for these models compared to normal mixture models.

# References

Aghaeepour, N., Finak, G., The FLOWCAP Consortium, et al. (2013). Critical assessment of automated flow cytometry analysis techniques. *Nature Methods* **10**, 228–238.

Anderson, E. (1935). The irises of the Gaspé Peninsula. *Bulletin of the American Iris Society* **59**, 2–5.

Arellano-Valle, R.B. and Genton, M.G. (2005). On fundamental skew distributions. *Journal of Multivariate Analysis* **96**, 93–116.

Azzalini, A. and Capitanio, A. (2003). Distributions generated by perturbation of symmetry with emphasis on a multivariate skew $t$ distribution. *Journal of the Royal Statistical Society B* **65**, 367–389.

Banfield, J.D. and Raftery, A.E. (1993). Model-based Gaussian and non-Gaussian clustering. *Biometrics* **49**, 803–821.

Böhning, D. (2000). Computer Assisted Analysis of Mixtures and Applications: Meta-Analysis, Disease Mapping, and Others. Chapman and Hall/CRC, London.

Branco, M.D. and Dey, D.K. (2001). A general class of multivariate skew-elliptical distributions. *Journal of Multivariate Analysis* **79**, 99–113.

Cabral, C.R.B., Lachos, V.H., and Prates, M.O. (2012). Multivariate mixture modeling using skew-normal independent distributions. *Computational Statistics and Data Analysis* **56**, 126–142.

Chen, W.C., Ostrouchov, G., Pugmire, D., P., et al. (2013). A parallel EM algorithm for model-based clustering applied to the exploration of large spatio-temporal data. *Technometrics* **55**, 513–523.

Everitt, B.S. and Hand, D.J. (1981). Finite Mixture Distributions. Chapman & Hall, London.

Fisher, R.A. (1936). The use of multiple measurements in taxonomic problems. *Annals of Eugenics* **7**, 179–188.

Forbes, F. and Wraith, D. (2014). A new family of multivariate heavy-tailed distributions with variable marginal amounts of tailweight: application to robust clustering. *Statistics and Computing* **24**, 971–984.

Fraley, C. and Raftery, A.E. (1998). How many clusters? which clustering methods? answers via model-based cluster analysis. *Computer Journal* **41**, 578–588.

Frühwirth-Schnatter, S. (2006). Finite Mixture and Markov Switching Models. Springer-Verlag, London.

Frühwirth-Schnatter, S. and Pyne, S. (2010). Bayesian inference for finite mixtures of univariate and multivariate skew-normal and skew-$t$ distributions. *Biostatistics* **11**, 317–336.

Guo, C., H., F., and Luk, W. (2012). A fully-pipelined Expectation-Maximization engine for Gaussian mixture models. In Proceedings of the International Conference on Field-Programmable Technology, 182–189.

Gupta, A.K. (2003). Multivariate skew-$t$ distribution. *Statistics* **37**, 359–363.

Hubert, L. and Arabie, P. (1985). Comparing partitions. *Journal of Classification* **2**, 193–218.

Kapoor, A., Hemani, H. anr Sakthivel, N., and Chaturvedi, S. (2015). MPI implementation of Expectation Maximization algorithm for Gaussian mixture models. In Proceedings of the 49th Annual Convention of the Computer Society of India, 517–523.

Karlis, D. and Santourian, A. (2009). Model-based clustering with non-elliptically contoured distributions. *Statistics and Computing* **19**, 73–83.

Kumar, N.S.L.P., Satoor, S., and Buck, I. (2009). Fast parallel Expectation Maximization for Gaussian mixture models on GPUs using CUDA. In Proceedings of the 11th IEEE International Conference on High Performance Computing and Communications, 103–109.

Kwedlo, W. (2014). A parallel EM algorithm for gaussian mixture models implemented on a NUMA system using OpenMP. In Proceedings of 22nd Euromicro International Conference on Parallel, Distributed, and Network-Based Processing, 292–298. Torino.

Lachos, V.H., Ghosh, P., and Arellano-Valle, R.B. (2010). Likelihood based inference for skew normal independent linear mixed models. *Statistica Sinica* **20**, 303–322.

Lee, S. and McLachlan, G.J. (2014). Finite mixtures of multivariate skew $t$-distributions: Some recent and new results. *Statistics and Computing* **24**, 181–202.

Lee, S.X. and McLachlan, G.J. (2013). On mixtures of skew-normal and skew $t$-distributions. *Advances in Data Analysis and Classification* **7**, 241–266.

Lee, S.X. and McLachlan, G.J. (2015). EMMIXcskew: an R package for the fitting of a mixture of canonical fundamental skew $t$-distributions. *arXiv:1509.02069* .

Lee, S.X. and McLachlan, G.J. (2016). Finite mixtures of canonical fundamental skew $t$-distributions: The unification of the restricted and unrestricted skew $t$-mixture models. *Statistics and Computing* **26**, 573–589.

Lin, T.I. (2010). Robust mixture modeling using multivariate skew-$t$ distribution. *Statistics and Computing* **20**, 343–356.

Lin, T.I., Ho, H.J., and Lee, C.R. (2014). Flexible mixture modelling using the multivariate skew-$t$-normal distribution. *Statistics and Computing* **24**, 531–546.

Lindsay, B.G. (1995). Mixture Models: Theory, Geometry, and Applications. NSF-CBMS Regional Conference Series in probability and Statistics, Vol. 5 (Institute of Mathematical Statistics and the American Statistical Association), Alexandria, VA.

McLachlan, G.J. and Basford, K.E. (1988). Mixture Models: Inference and Applications. Marcel Dekker, New York.

McLachlan, G.J. and Lee, S.X. (2016). Comment on "On nomenclature for, and the relative merits of, two formulations of skew distributions" by A. Azzalini, R. Browne, M. Genton, and P. McNicholas. *Statistics and Probaility Letters* **116**, 1–5.

McLachlan, G.J. and Peel, D. (2000). Finite Mixture Models. Wiley, New York.

Mengersen, K.L., Robert, C.P., and Titterington, D.M. (2011). Mixtures: Estimation and Applications. John Wiley & Sons, New York.

Pyne, S., Hu, X., Wang, K., et al. (2009). Automated high-dimensional flow cytometric data analysis. *Proceedings of the National Academy of Sciences USA* **106**, 8519–8524.

R Core Team (2016). R: A Language and Environment for Statistical Computing. R Foundation for Statistical Computing, Vienna, Austria. ISBN 3-900051-07-0.

Sahu, S.K., Dey, D.K., and Branco, M.D. (2003). A new class of multivariate skew distributions with applications to bayesian regression models. *The Canadian Journal of Statistics* **31**, 129–150.

Titterington, D.M., Smith, A.F.M., and Markov, U.E. (1985). Statistical Analysis of Finite Mixture Distributions. Wiley, New York.

Wang, K., Ng, S.K., and McLachlan, G.J. (2009). Multivariate skew $t$ mixture models: applications to fluorescence-activated cell sorting data. In H. Shi, Y. Zhang, M.J. Bottema, B.C. Lovell, and A.J. Maeder (Eds), Proceedings of Conference of Digital Image Computing: Techniques and Applications, 526–531. Los Alamitos, California.

Wraith, D. and Forbes, F. (2015). Location and scale mixtures of Gaussians with flexible tail behaviour: properties, inference and application to multivariate clustering. *Computational Statistics and Data Analysis* **90**, 61–73.